# Sub-zeptojoule detection of terahertz pulses by parametric frequency upconversion

Défi Junior Jubgang Fandio,[1] Aswin Vishnuradhan,[1] Eeswar Kumar Yalavarthi,[1] Wei Cui,[1] Nicolas Couture,[1] Angela Gamouras,[1,2] and Jean-Michel Ménard[1,2]

[1]Department of Physics, University of Ottawa, Ottawa, Ontario K1N 6N5, Canada
[2]National Research Council Canada, Ottawa, Ontario K1A 0R6, Canada

*Corresponding author: jean-michel.menard@uottawa.ca

We combine parametric frequency upconversion with single-photon counting technology to achieve detection sensitivity down to the terahertz (THz) single-photon level. Our relatively simple detection scheme employs a near-infrared ultrafast source, a GaP nonlinear crystal, optical filters, and a single photon avalanche diode. This configuration is capable of detecting a weak THz signal with an energy of 590 zJ contained within a single pulse. Through averaging over 50k data points, the configuration can resolve a 0.5 zJ pulse energy, corresponding to an average of 0.5 photon per pulse. The corresponding noise-equivalent power and THz-to-NIR photon detection efficiency are $4.1 \times 10^{-17}$ W/Hz$^{1/2}$ and 0.19%, respectively. To test our scheme, we perform spectroscopy of water vapor between 1.0 and 3.7 THz and obtain results in agreement with to those acquired with a standard electro-optic sampling (EOS) method. Our technique provides a 0.2 THz spectral resolution offering a fast alternative to EOS THz detection for monitoring specific spectral components in THz spectroscopy, imaging and communications applications.

Terahertz (THz) technology has attracted interest in a plethora of applications including non-invasive quality control of pharmaceuticals [1,2], wireless communication [3,4], bio-imaging [5], and the inspection of biohazards and explosives for security purposes [2,3,6]. The THz band has also been exploited to investigate advanced materials like 2D structures, semiconductor nanostructures and quantum materials [7–10]. THz systems used for these applications are still relatively limited in their performance due to technical difficulties in the efficient detection and characterization of THz radiation. As a result, many research groups have focused on improving the performance of THz detectors in terms of their spectral resolution, acquisition speed and, most particularly, their sensitivity [11,12]. The latter is directly related to the detection efficiency, which is also a critical property to enable quantum applications taking advantage of THz radiation's unique properties, notably used in spectroscopy, imaging and wireless communication [13–15]. Ultimately, quantum applications require single THz photon sensitivity. Although single-photon detection of visible and near-infrared (NIR) photons can now be routinely achieved with semiconductor-based photodetectors, those devices are inefficient in the THz range. Experimentalists must instead rely on special thermal detectors, notably based on superconductors and cryogenic equipment [16,17]. Other technologies such as Golay cells, microresonators and pyroelectric detectors are intrinsically limited in sensitivity by surrounding thermal noise. In parallel, electro-optical sampling (EOS) is also an extremely sensitive detection technique but this technique typically requires the collection of multiple data points to fully reconstruct a THz waveform, which inherently prevents the resolution of single photons [18]. Parametric upconversion of THz photons in the NIR region has been implemented as an alternative to achieve fast and sensitive detection of THz radiation using broadly-used room-temperature photodetectors. Previous demonstrations relying on different nonlinear media, such as GaAs [19,20], GaSe [21], KTP [22], DAST [23,24], and LiNbO$_3$ [25–27], have shown that the upconverted NIR signal preserves the spatial, temporal, and spectral information of the THz input signal [28]. Using this technique, a minimum detectable THz energy per pulse of 130 zJ was demonstrated in LiNbO$_3$ at a frequency of 1 THz [25]. In this Letter, we present the detection of sub-zeptojoule pulsed THz radiation through parametric frequency upconversion in gallium phosphide (GaP) from 1 to 3.7 THz. Our technique combines a set of polarization and spectral filters, a monochromator, and a NIR single-photon counting module (SPCM), all operating at room temperature. At 2 THz, we demonstrate a minimum detectable THz energy per pulse of 0.5 zJ after averaging over 50k pulses, which corresponds to a noise equivalent power (NEP) of $4.1 \times 10^{-17}$ W/Hz$^{1/2}$.

Figure 1 shows a schematic depiction of the experimental setup. It consists of an amplified laser source delivering 180 fs pulses at a central wavelength of 1035 nm (289.86 THz), with a pulse repetition rate set to 50 kHz. The laser output beam is divided into two arms by a beam splitter (BS). The transmitted beam (10% of the incident 6 W power) is directed toward a 300 µm-thick 110-oriented GaP crystal for broadband THz generation through optical rectification. A half-wave plate ($\lambda/2$) and a polarizer ($P_1$) are placed along the generation beam path to adjust the incident optical power on the generation crystal and, consequently, control the generated THz power. The THz pulse energy is calculated from the calibrated time-resolved signal using:

$$E_{pulse} = \int P(t)dt = \int \left(\pi r_{THz}^2 \times \epsilon_0\, c\, |\varepsilon_{peak}\, \varepsilon(t)|^2\right) dt \quad (1)$$

where $P(t)$ is the time-resolved THz power obtained, $\varepsilon(t)$ is the normalized THz field transient, $\varepsilon_{peak}$ is the THz peak field measured from the EOS trace [29], and $r_{THz}$ = 220 µm (1/e$^2$) is the measured THz beam radius using the knife edge method.

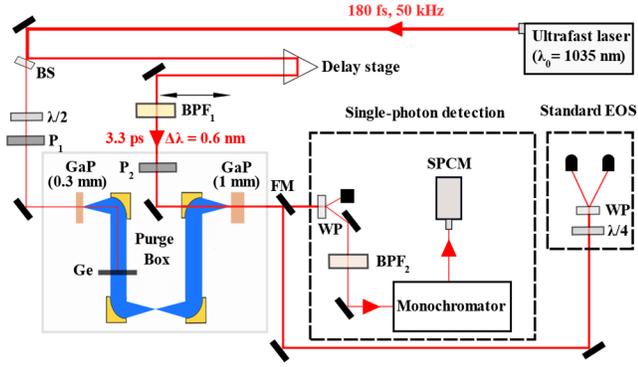

**Fig. 1.** Experimental setup of the THz generation and detection techniques. BS-Beam splitter (10/90), Polarizers-$P_1$ & $P_2$, BPF-bandpass filter (BPF$_1$: $\lambda_0$ = 1030 nm, $\Delta\lambda$ = 0.6 nm, BPF$_2$: $\lambda_0$ = 1022 nm, $\Delta\lambda$ = 10 nm), $\lambda/2$-Half-wave plate, $\lambda/4$-Quarter-wave plate, FM- flip mirror, WP-Wollaston prism, SCPM-Single-photon counting module.

The beam reflected from the BS (90% of the incident 6 W power) serves as the gating beam for THz detection. It is first transmitted through a bandpass filter BPF$_1$, resulting in a spectrum centered at 1028.2 nm ($v_0$ = 291.77 THz) with a 0.6 nm (or 0.17 THz) spectral linewidth (FWHM). Consequently, the NIR pulse amplitude is then stretched to a 4.7 ps duration (FWHM). A polarizer ($P_2$) is then used to ensure a horizontal polarization state. The gating pulse is overlapped in time and space with the generated THz radiation inside a 1 mm-thick 110-oriented GaP crystal to achieve parametric frequency conversion.

After the nonlinear crystal, the Wollaston prism (WP) is used to preferentially select vertically aligned polarization components corresponding to NIR photons that nonlinearly interacted with the THz pulse via type I phase-matching. This step reduces the background NIR signal by a factor of 5000. The bandpass filter BPF$_2$ centered at 1022 nm ($v_0$ = 293.54 THz) preferentially transmits up-converted THz photons while reducing the residual gating pulse power by about four orders of magnitude at 1030 nm. Finally, the up-converted photons are detected using a monochromator (iHR320) and a SPCM (SPCM-CD-34-62-H). Standard EOS measurements are performed for calibration and comparison purposes by removing BPF$_1$ and redirecting the gating beam to a quarter waveplate ($\lambda/4$), a Wollaston prism and a balanced detection scheme as indicated in Fig. 1. The THz beam path is fully enclosed in a box, which can be purged with dry air to achieve <2 % relative humidity.

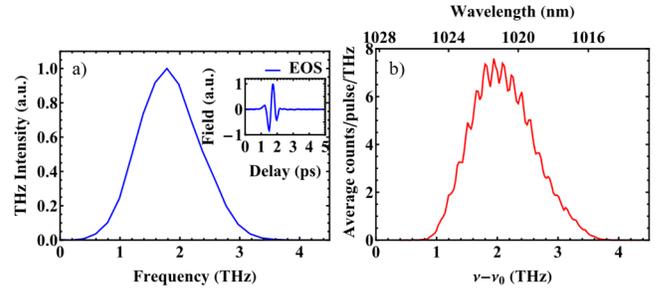

**Fig. 2.** (a) THz spectrum calculated from the Fourier transform of the THz field transient measured via EOS (inset). (b) THz field spectrum detected via THz upconversion.

A comparison of the THz spectra recorded with EOS and our single-photon detection configuration is presented in Fig. 2. Parameters in both experiments are set to obtain the same spectral resolution of 0.2 THz. For the EOS technique, this resolution is achieved by acquiring the time-domain signal over a 5 ps temporal window, while, for the single-photon technique, the entrance and exit slits of the monochromator are adjusted to achieve a 0.125 nm resolution of the upconverted NIR signal. Note that the duration of the gating pulse in the single-photon configuration also contributes to determine the THz spectral resolution. Both configurations yield similar spectra characterized by a Gaussian-like shape peaking around 2 THz with a width ranging from ~1.0 to 3.5 THz. In Fig. 2b, the spectral intensity obtained from monitoring the upconverted NIR photons is slightly lower around 1 THz in comparison to the EOS results. This dissimilarity is caused by the response function of the filter BPF$_2$ used to decrease the unmodulated gating pulse. This filter also induces periodic amplitude oscillations, mostly visible around the peak frequency, which was not found to affect spectroscopy measurements.

The acquisition time for every THz frequency component of the upconverted signal is determined by the SPCM and frequency counter used for electronic read-out. In our experiment, this unit can acquire up to 75,000 counts per second. Therefore, the single-photon detection scheme can be advantageous over EOS to monitor a specific spectral component since it intrinsically requires fewer data points and can therefore be performed on a faster time scale. In our experiment, the THz spectrum shown in Fig. 2a is measured in 115 s, while each spectral component in Fig. 2b is acquired in 1s, and both display a similar signal-to-noise ratio.

We explore our detector's response at different incident THz pulse energies. Figure 3a shows the sensitivity of the upconversion detection system using different SPCM integration times of 0.01 s, 0.1 s and 1 s. In this study, the monochromator is set to resolve frequencies centered at 2 THz with a 1.3 THz bandwidth to cover the entire generated THz spectrum. The THz signal is expressed as counts/pulse and is calculated from the total number of counts provided by the SPCM, divided by the laser repetition rate and the integration time. The minimum detectable THz energy is reached when the signal is equal to the standard deviation, which corresponds to a signal to noise ratio of 1. Using this metric, our configuration can resolve a THz pulse with 0.5 zJ energy using a SPCM integration time of 1 s. Such a pulse has an average of 0.5 photon/pulse at 2 THz. Our SPCM monitors 315 counts from the 50,000 pulses (or ~0.006 counts/pulse) corresponding to a 0.19% quantum detection efficiency. Reducing the integration time to 10 ms, while keeping all other parameters constant, increases the noise and, consequently, the minimum detectable THz pulse energy to 4.5 zJ. The minimum sensitivity of 0.5 zJ shown here is two orders of magnitude weaker than values previously measured with other schemes relying on different nonlinear materials such as GaAs [19] and LiNbO$_3$ [25] for parametric conversion.

Generally, a room-temperature THz detector reaching this sensitivity level would provide readings limited by thermal background photons. However, in our configuration, we are only sensitive to THz photons overlapping in space and time with the ps NIR gating pulse. Also, these photons must impinge on the crystal within a small acceptance angle to satisfy phase matching conditions. Finally, the nonlinear crystal used for upconversion is transparent to THz radiation, which, according to Kirchhoff's law, limits the emission of parasitic thermal THz photons. Further precautions could be taken to decrease the density of thermal photons, but our results do not indicate a relevant contribution from this black body radiation.

Fitting the data in Fig. 3a with a linear function yields a detection efficiency of $\eta_{THz \to NIR}$ (2 THz) = 0.19 %, which is, to the best of our knowledge, the largest THz-to-NIR detection efficiency achieved so far in an upconversion configuration. Previous THz parametric detectors have reported photon detection efficiencies of 7.5 × 10$^{-4}$ % in GaP [30], 5.9 × 10$^{-4}$ % in GaAs [20], and 8.4 × 10$^{-5}$ % in GaSe [21]. Saturation is observed at pulse energy exceeding 400 zJ since the SPCM record one count even if two photons are simultaneously reaching the detector. The conversion efficiency of our THz detection technique can still be optimized by using a different nonlinear crystal with a higher second-order nonlinear coefficient, such as GaSe or DAST. Furthermore, a gating beam with a wavelength in the visible region would allow the photodiode to operate in a regime of higher quantum detection efficiency, directly leading to a higher THz detection sensitivity.

NEP is a crucial parameter to attest of the performance of an optical detector. In our detection scheme, NEP is evaluated by considering the specifications of the SPCM and the detection efficiency of the upconverted photons using this expression [20]:

$$NEP_{THz} = \left(\frac{\nu_{THz}}{\nu_{up} \times \eta_{THz \to NIR}}\right) \times h\,\nu_{up} \frac{\sqrt{DCR}}{\eta_{SPCM}}, \qquad (2)$$

where $\nu_{THz}$ = 2 THz is the THz frequency, $\nu_{up}$ = 293.71 THz is the upconverted radiation frequency, $\eta_{THz \to NIR}$ is the detection efficiency of the entire system, $DCR$ and $\eta_{SPCM}$ represent the SPCM dark-count rate and photodetector efficiency, respectively. Considering the measured $DCR$ of 30 counts/s and the SPCM photodetection efficiency $\eta_{SPCM}$ at 1022 nm of 9.4 %, we obtain a NEP$_{THz}$ of 4.1×10$^{-17}$ W/Hz$^{1/2}$. This value correspond to an improvement of 5 orders of magnitude over THz detection schemes reported in previous work [19,22].

Figure 3b shows the detected counts per pulses as a function of the total number of incident THz pulses. Different incident pulse energies are explored from 3.3 to 2370 zJ. As expected, the number of counts/pulses remains constant irrespective of the SPCM integration time. However, the error, corresponding to the standard deviation, increases as we reduce the number of incident pulses. These results show that our THz detection scheme can resolve THz pulses as weak as 39 zJ from averaging over 10 pulses, which can be achieved in 0.2 ms. Based on the pulse energy and integration time study presented in Fig. 3b, our detection scheme can resolve a single count per individual pulse, with 80% probability, when the input THz pulse has an energy of 590 zJ. The detection probability increases to 97 %, at 2370 zJ.

In a last experiment, we demonstrate the spectral accuracy of our configuration by performing spectroscopy of ambient air. The THz pulse energy was fixed to 350 zJ and the relative humidity in ambient air was kept constant at 30 ± 5 % over a ~39 cm THz beam propagation length. Figure 4 compares the transmission spectrum obtained with our upconversion detection scheme (red) and the EOS technique (blue) using a spectral resolution of 0.2 THz. Both spectra are in excellent agreement with each other, and are also consistent with the water vapor transmission spectrum reported by HITRAN [31]. Although the spectroscopy performed in this work is limited within the 1.0 - 3.7 THz window, it would be possible to shift this detection window in our scheme by using a different nonlinear crystal than GaP with absorption and phase matching conditions allowing efficient upconversion at different THz frequencies.

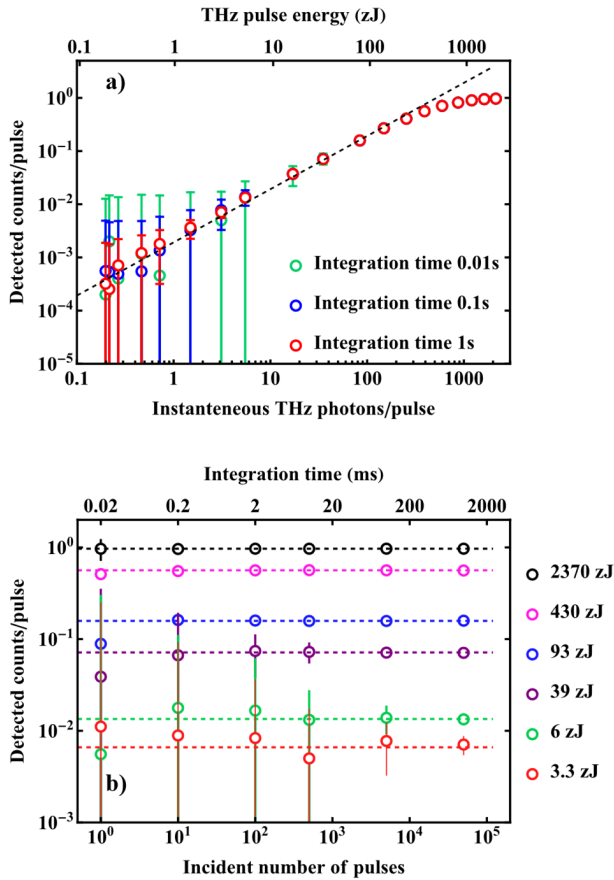

**Fig. 3.** (a) Sensitivity at 2 THz (1021.4 nm). The SPCM integration time is set to 1 s, 0.1 s and 0.01 s. The minimum sensitivity of detection is 0.5 zJ/pulse. (b) Detected counts/pulse as a function of incident THz pulses.

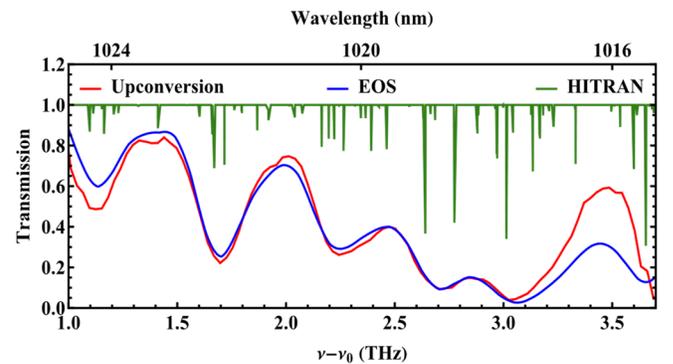

**Fig. 4.** The ambient air transmission spectrum measured with our THz detection scheme shows a good agreement with the spectrum obtained from THz-TDS with the same humidity. Both spectra are consistent with HITRAN[31].

In conclusion, we demonstrate a fast and highly-sensitive room-temperature THz detection scheme based on parametric frequency upconversion and single-photon counting technology. This technique can directly map every spectral component of the THz field amplitude within a 4.7 ps temporal window of the spectrally filtered gating beam pulse width. We show that this technique can be orders of magnitude faster than standard EOS in resolving single spectral components of the THz field. Although we use a spectral resolution of 0.2 THz in our experiments, this resolution can be improved by using a gating pulse with longer pulse duration as it has been recently demonstrated [22,25]. The minimum detectable THz pulse energy of our system is 0.5 zJ at 2 THz after integrating over 50k pulses. This sensitivity represents a detection NEP of $4.1 \times 10^{-17}$ W/Hz$^{1/2}$ and an overall system detection efficiency of 0.19%. We also show that a single THz pulse with a 2370 zJ energy can be resolved with a 97% probability. Such high sensitivity may enable long-distance sixth-generation wireless communication systems relying on frequencies above 100 GHz [32]. Finally, we demonstrate that our configuration can be used for THz spectroscopy within the 1.0 - 3.7 THz spectral range, although the real advantage of this technique resides in the fast and sensitive monitoring of a specific spectral component. We believe this configuration will allow new applications requiring the resolution of very weak THz components and lead to THz quantum applications without the need for any cryogenic equipment.

**Acknowledgments.** We acknowledge funding from the High Throughput and Secure Networks Challenge Program at the National Research Council of Canada (HTSN-702), the NSERC Discovery funding program (RGPIN-2016-04797, RGPIN-2023-05365), the Canada Foundation for Innovation (CFI) (Project Number 35269), and the National Research Council of Canada via the Joint Centre for Extreme Photonics (JCEP).
**Disclosures.** The authors declare no conflicts of interest.
**Data Availability** Data underlying the results presented in this paper are not publicly available at this time but may be obtained from the authors upon reasonable request.